\def\be{\begin{equation}}
	\def\ee{\end{equation}}
\def\ber{\begin{eqnarray}}
	\def\eer{\end{eqnarray}}
\def\bers{\begin{eqnarray*}}
	\def\eers{\end{eqnarray*}}

\RequirePackage{latexrelease}
\documentclass[aps,prb,twocolumn,showpacs,groupedaddress,amsmath,amssymb]{revtex4-1}
\usepackage[colorlinks=true,citecolor=black,linkcolor=blue,urlcolor=blue]{hyperref}

\usepackage{amsmath}
\usepackage{titlesec}
\titlespacing*{\section}{0pt}{0.8\baselineskip}{0.6\baselineskip}

\usepackage{latexrelease}
\usepackage{latexsym}
\usepackage{xcolor}
\usepackage{float}
\usepackage{dcolumn}   
\usepackage{amsmath}
\usepackage{indentfirst}
\usepackage{multirow}
\usepackage{graphicx}    
\usepackage{subfigure}
\usepackage{comment}
\usepackage{color}
\usepackage{makecell}
\usepackage{longtable}
\usepackage{natbib}
\usepackage{ifthen}
\newboolean{includefigs}
\setboolean{includefigs}{true}      
\newboolean{includetext}
\setboolean{includetext}{true}     
\newcommand{\condcomment}[2]{\ifthenelse{#1}{#2}{}}
\usepackage[utf8]{inputenc}
\usepackage{epstopdf}
\usepackage{wrapfig}
\usepackage{hyperref}
\DeclareUnicodeCharacter{0308}{XXXXXXXXXXX}

\begin{document}
	
	\title{Biphenylene Nanoribbon as Promising Electrocatalyst for Hydrogen Evolution}

\author{Radha N Somaiya}
\affiliation{Materials Modeling Laboratory, Department of Physics, IIT Bombay, Powai, Mumbai 400076, India}

\author{Zicong Marvin Wong}
\affiliation{Institute of High Performance Computing, Agency for Science, Technology and Research, 1 Fusionopolis Way, \#16-16 Connexis, Singapore 138632, Singapore}

\author{Brahmananda Chakraborty}
\affiliation{Homi Bhabha National Institute, Trombay, Mumbai-400085, India}
\affiliation{High Pressure \& Synchrotron Radiation Physics Division, Bhabha Atomic Research Centre, Trombay, Mumbai-400085, India}

\author{Teck Leong Tan}
\affiliation{Institute of High Performance Computing, Agency for Science, Technology and Research, 1 Fusionopolis Way, \#16-16 Connexis, Singapore 138632, Singapore}

\author{Aftab Alam}
\email{aftab@iitb.ac.in}
\affiliation{Materials Modeling Laboratory, Department of Physics, IIT Bombay, Powai, Mumbai 400076, India}

\date{\today}%
	

\begin{abstract}
		Designing efficient, metal free, and in-expensive catalyst for electrochemical hydrogen evolution reaction (HER) is crucial for large scale clean and green energy production. Recently synthesized 1D Biphenylene nanoribbons (BPRs) display few exciting properties originating from the unique co-existence of 4, 6 and 8 coordinated carbon rings. Here, we present a first principles calculation of the electronic structure and electrocatalytic activity of various sized N-BPRs (N indicates the width). The electrocatalytic performance is evaluated based on several descriptors including electronic property, carrier mobility, Gibbs free energy ($\Delta$G$_{HER}$), exchange current density etc. The electronic properties are crucially sensitive to the width (N) of BPRs transiting it from semiconducting to metallic nature at N=18. The p$_z$ orbitals of C-atoms from the central tetragonal rings are mainly responsible for the decrease in band gap with increasing width. The room temperature electron mobility is found to be as high as $\sim6.3\times$10$^4$ cm$^2$V$^{-1}$s$^{-1}$, while hole mobility is relatively lower in magnitude. Gibbs free energy also depends sensitively on the width of BPRs as evident from the p-band center analysis. We propose 15-BPR as the most promising candidate for electrocatalytic activity with extremely small overpotential ($\approx$0.005 V) and a high exchange current density, much better than the state-of-the-art Pt(111). A close inspection of the elementary reactions of HER (Volmer, Heyrovsky and Tafel) confirms Volmer-Tafel mechanism to be most dominant on 15-BPR with Tafel as the rate-determining step with a barrier of 0.56 eV. The present study provides a deeper insight into the excellent HER catalytic activity of a newly synthesized BPR which is inexpensive and is expected to hasten experimental validation towards H$_2$ production.  
\end{abstract}

\maketitle
	
	
	\section{Introduction}
	Hydrogen (H$_2$) being one of the cleanest form of energy is generated from water using appropriate decomposition technology. It has several advantages including no greenhouse gas emission, high energy density, zero emission source etc.\cite{turner2004sustainable}. H$_2$ has received global attention because of the growing demands of energy and increased environmental pollution due to the ever-increasing world population. Of the various approaches available for the generation of H$_2$, photoelectrochemical water splitting based hydrogen evolution reaction (HER) is considered as an eco-friendly and viable approach\cite{wang2021borophene}. Hydrogen based economy is a blooming industry which has opened up a new era of renewable energy commercialization, including fuel cell cars with zero emission as a promising example\cite{uosaki2016highly}.
	
	Ever since, electrochemical HER has been widely studied both from theoretical as well as experimental front\cite{gurney1931quantum,parsons1958rate,kita1966periodic}. Choice of efficient catalysts to drive HER for H$_2$ production is critical as they play a crucial role in accelerating the proton-electron pairs to form H$_2$ at very low overpotentials. Platinum (Pt) based catalysts display outstanding HER performance amongst all the available catalysts with small Tafel slopes and a high exchange current density. But the high cost and scarcity of Pt restricts its widespread use\cite{bai2021recent, skulason2010modeling}. For sustainable H$_2$ production, the development of stable, cost efficient and readily available catalysts with high catalytic activity are highly desirable. To meet the requirement and replace Pt, a variety of inexpensive catalysts have been proposed indicating better activity, such as transition metal oxides, sulphides, carbides, selenides, nitrides, and phosphides. However, these materials rely on metal-H bond which suffers corrosion under extreme environmental conditions\cite{zhang2016active}. An efficient HER catalyst requires effective electron transfer at the surface as well as within the catalyst itself. But these criteria remain discontented due to poor conductivity by most non-metallic electrocatalysts like MoS$_2$. One strategy to circumvent this bottle neck is to integrate conducting component such as carbon into the electrocatalyst\cite{murthy2018recent}. The above idea is selectively highlighted in few reports, for instance, MoS$_2$ nanoparticles supported on highly conducting and inert carbon\cite{hinnemann2005biomimetic}. Carbon and its role as an HER catalyst is an intensive topic of research from the start of U.S. space program in 1960s\cite{das2014extraordinary}.

	Carbon, an abundant element on the earth, exhibits various allotropes owing to its flexibility to hybridize differently. Naturally existing forms are graphite and diamond having \textit{sp$^2$} and \textit{sp$^3$} hybridization, respectively. Graphene is the first two-dimensional (2D) allotrope, which acquire robust stability, a semimetal with zero band gap and high carrier mobility (2$\times$10$^5$ cm$^2$V$^{-1}$s$^{-1}$)\cite{geim2007rise}. It has gained enormous attention owing to its outstanding properties useful for a broad spectrum of applications. However it possess a high value of Gibbs free energy ($\Delta$G$_{HER}$), a catalytic descriptor determining the strength of H adsorption (smaller the $\Delta$G$_{HER}$, better the HER activity), indicating its poor catalytic performance as it is chemically inert. This can be further enhanced via (i) intrinsic defects such as vacancies, interstitial (ii) doping with foreign atoms (iii) functionalization etc.\cite{zheng2014toward}. This then motivated a search for other low-dimensional materials leading to experimental synthesis of graphdiyne ($\Delta$G$_{HER}$=1.18 eV) in 2010\cite{li2010architecture,hui2019efficient} and T-carbon in 2017\cite{zhang2017pseudo}. Further careful efforts lead to the discovery of yet other types of carbon allotropes, both experimentally and theoretically, which includes penta-graphene ($\Delta$G$_{HER}$=2.72 eV), holey graphyne ($\Delta$G$_{HER}$=0.40 eV), C$_2$N  ($\Delta$G$_{HER}$=-0.70 eV), g-C$_3$N$_4$ ($\Delta$G$_{HER}$=-0.54 eV) and N-graphene ($\Delta$G$_{HER}$=0.57 eV), g-CN ($\Delta$G$_{HER}$=-0.40 eV), carbon nanotubes ($\Delta$G$_{HER}$=1.15 eV), graphene nanocages ($\Delta$G$_{HER}$=0.032 eV), carbon cloth ($\Delta$G$_{HER}$=$\sim$0.13-0.29 eV), etc.\cite{chodvadiya2023transition,zhang2018transition,holmberg2015ab,zheng2014hydrogen,hao2021defect,chen2021transition,wei2020nanoribbon}. Similar to graphene, other pristine carbon allotropes also display poor catalytic performance which may arise from their non-conductive nature and structural morphology. Interestingly, Graphene nanocages acquire very small value of $\Delta$G$_{HER}$ indicating its potential as an HER catalyst but its experimental synthesis is rather challenging\cite{wei2020nanoribbon, rasool2022insight}.

	Among the various planar \textit{sp$^2$} hybridized carbon allotropes, only graphene is realized to purely crystallize in hexagonal network. Limited efforts have been observed to realize three-coordinated carbon networks consisting of non-hexagonal carbon rings. In 1968, Biphenylene (BPH) was first predicted to show \textit{sp$^2$} hybridization with alternating four, six, and eight carbon rings\cite{balaban1968chemical}. In 2014, a method to octafunctionalize BPH was suggested from the molecular precursors of isomeric graphene nanostructures\cite{schlutter2014octafunctionalized}. Very recently, using a two-step dehydrofluorination polymerization method, a one-dimensional (1D) counterpart possessing \textit{sp$^2$} hybridized biphenylene network was successfully synthesized. The biphenylene nanoribbons (BPRs) with varying widths ranging from 12 to 42 carbon atoms were synthesized. A rapid decrease in the electronic band gap from 2.35 to 0 eV was observed as the width increases, indicating onset of metallic characteristics as one approaches from 1D nanoribbon to 2D BPH. The experimental findings were further supported by theoretical calculations, which also confirmed the decreasing trend of the band gap with an increase in BPR size\cite{fan2021biphenylene,shen2022electronic}. Similar to graphene nanoribbons\cite{han2007energy,son2006energy}, 1D quantum confinement opens energy band gap in BPRs making them attractive for wide range of applications. 2D BPH is observed to be metallic with an n-type Dirac cone at about 0.64 eV above the Fermi level (E$_F$) and is found to exhibit many interesting properties. Its unique geometrical porous structure makes it a better HER candidate ($\Delta$G$_{HER}$=0.29 eV) compared to graphene ($\Delta$G$_{HER}$=1.41 eV)\cite{ge2016theoretical,ferguson2017biphenylene,luo2021first,han2022biphenylene,sahoo2023activation}. This has motivated us to study and gain an in-depth understanding of the catalytic mechanism of its 1D counterpart nanoribbons, the so-called BPR.

	In this manuscript, our focus lies in a microscopic understanding of the electronic structure, carrier transport and HER mechanism of various BPRs (with different widths) employing density functional theory (DFT) and the transition state calculations. The carrier transport, including the effect of prominent scattering on deformation potential and mobility, is also studied using the Boltzmann transport equation and the deformation potential theory. BPRs of varying widths from 12 to 42 carbon atoms were studied. Of these, 12-BPR, 15-BPR, 18-BPR, and 21-BPR (N-BPR implies BPR with 2N carbon atoms) are found suitable as HER catalyst with small values of Gibbs free energies. Specifically, 15-BPR shows the best potential for HER activity with a small hydrogen adsorption Gibbs free energy ($\Delta$G$_{HER}$) of 0.005 eV, which is even much smaller compared to the state-of-the-art catalyst Pt(111) (-0.09 eV). This is further affirmed by the analysis of the simulated volcano plot and the p-band center catalytic descriptor. Moreover, we also investigated the HER mechanism of smallest 6-BPR and 15-BPR to understand the evolution mechanism of H$_2$ molecule.

	
	
	\section{Computational Details}
	
	The First-principles calculations were carried out using DFT\cite{hohenberg1964inhomogeneous} by employing projector augmented wave potentials\cite{blochl1994projector}, as implemented within Vienna ab initio simulation package (VASP) code\cite{kresse1996efficient,kresse1999ultrasoft}. Perdew-Burke-Ernzerhof (PBE)\cite{perdew1996generalized} within the Generalized gradient approach (GGA)\cite{perdew1986accurate} was used to incorporate the exchange and correlation functional. We have utilized a kinetic energy cutoff of 550 eV and a vacuum of 30 \AA\ is added along the y- and z-directions to minimize any spurious interactions introduced by the periodic boundary conditions. Heyd-Scuseria-Ernzerhof (HSE06)\cite{heyd2003hybrid} screened hybrid functional is adopted to better estimate the electronic band gaps. Grimme's DFT-D3\cite{klimevs2011van,dion2004van} dispersion correction is implemented to account for the weak van der Waals interactions. The Brillouin zone is integrated using $\Gamma$-centered Monkhorst-Pack\cite{monkhorst1976special,pack1977special} 10$\times$1$\times$1 and 15$\times$1$\times$1 k-meshes to investigate the structural and electronic properties, respectively. The energy and force convergence criteria are set at $1$x$10^{-7}$ eV and -0.001 eV/\AA\, respectively. Bader charge analysis is carried out to investigate the charge transfer mechanism using the method developed by Henkelman\cite{bader1985atoms,henkelman2006fast}. Additional computational details are given in the Supporting information\cite{ESI}; also see Ref. \cite{blochl1994improved}.
		
	
	\section{Results and Discussion}

 \begin{figure}[t]
	\centering
	\includegraphics[width=\columnwidth]{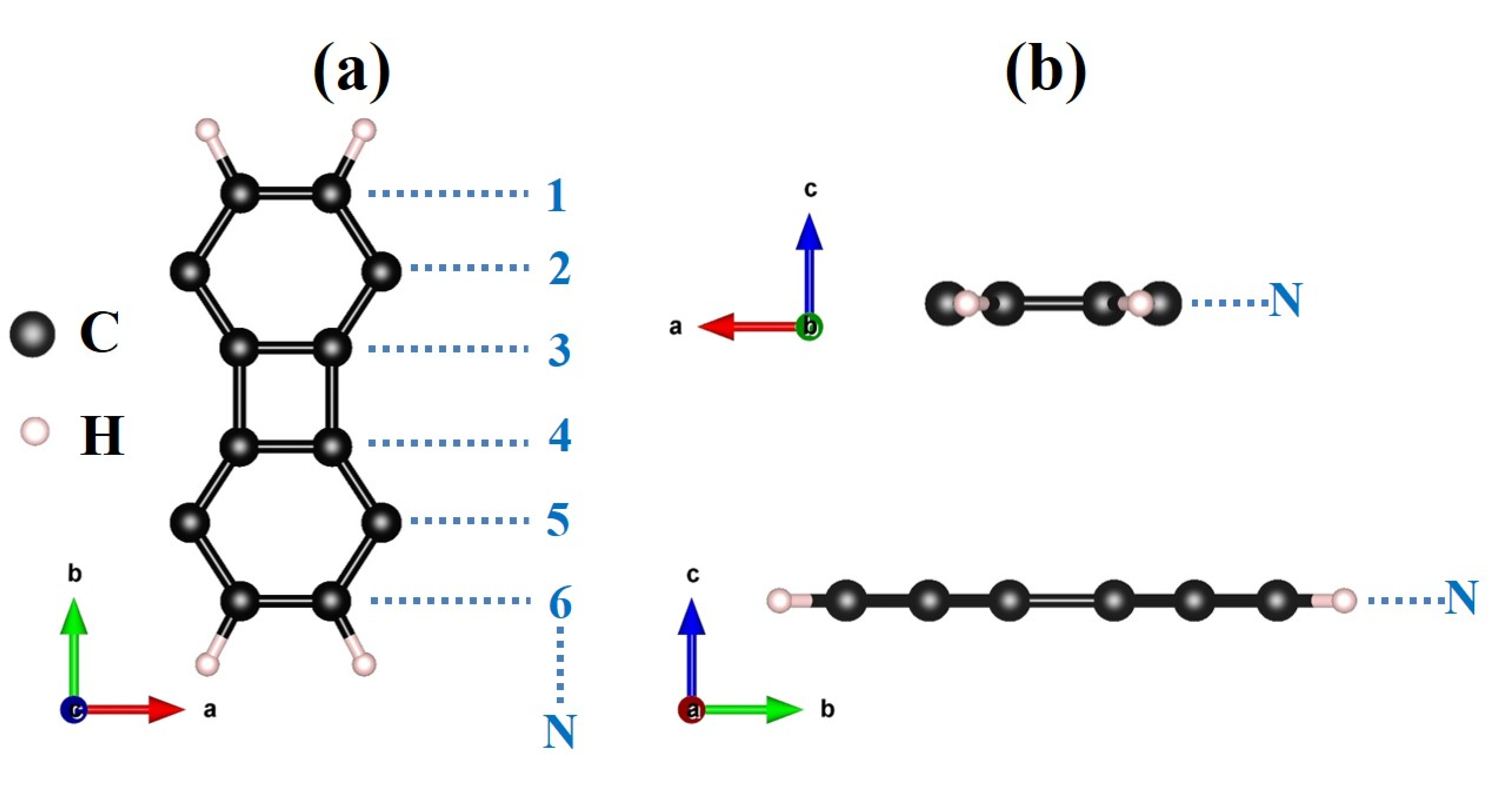}
	\caption{(a) Top and (b) side views of Biphenylene Nanoribbon (N-BPR). N refers to the width of BPRs having 2N carbon atoms in the unit cell. BPRs with width N = 6, 9, 12, 15, 18, and 21 are studied here. The black and pink solid spheres represent carbon and hydrogen atoms, respectively.} 
	\label{Fig. 1}
\end{figure}

\begin{table}[b]%
	\caption{Lattice constant, BPR width, measured (Expt.) and simulated electronic band gap calculated using PBE and HSE06 functionals. D (ID) indicates the direct (indirect) nature of band gap while M indicates metallic nature of nanoribbon.}
	\begin{center}
		\begin{ruledtabular}
			\begin{tabular}{ccccccc}
				N-BPR & a & width & E$_g$ & E$_g$ & E$_g$ \\ [0.5ex]
				&  &  & (PBE) & (HSE06) & (Expt.) \\ [0.5ex]
				& (\AA) & (\AA) & (eV) & (eV) & (eV) \\ [0.5ex] \hline
				6-BPR & 4.389  & 8.158 & 1.03 (D) & 1.62 (D) & 2.35 (D) \\ [0.5ex]
				9-BPR & 4.406  & 11.984 & 0.49 (ID) & 1.00 (ID) & 1.44 (ID) \\ [0.5ex]
				12-BPR & 4.416  & 15.806 & 0.03 (ID) & 0.42 (ID) & 0.84 (ID) \\ [0.5ex]
				15-BPR & 4.422  & 19.610 & M & 0.25 (ID) & 0.70 (ID) \\ [0.5ex]
				18-BPR & 4.427  & 23.341 & M & M & 0.36 (ID) \\ [0.5ex]				
				21-BPR & 4.434  & 27.125 & M & M & M \\ [0.5ex]				
			\end{tabular}
		\end{ruledtabular}
	\end{center}
	\label{table:table1}
\end{table}

	Biphenylene is a planar non-hexagonal allotrope of carbon made of 4-6-8 membered carbon rings. Each carbon atom connects with three other carbon atoms possessing \textit{sp$^2$} hybridization. The BPRs are systematically described based on their widths (N). Figure ~\ref{Fig. 1} depicts the optimized structure of N-BPR which is periodic along the x-direction and has a fixed width along the y-direction. The experimentally synthesized smallest BPR has a width of 6 carbon (C) atoms and hence abbreviated as 6-BPR. It can also be viewed as one biphenylene unit, which comprises of a total of 12 C-atoms with the ends passivated by four hydrogen (H) atoms. Subsequent addition of hexagonal rings in 6-BPR forms 9-, 12-, 15-, 18-, and 21-BPR with an increment of 6 C-atoms in the unit cell each time. Table ~\ref{table:table1} indicates the optimized structural properties for N-BPRs (N = 6, 9, 12, 15, 18, and 21) which matches fairly well with the experimental and other theoretical outcomes\cite{fan2021biphenylene,shen2022electronic}. Since the structure is composed of 4-6-8 membered carbon rings, the C-C bond distance varies between 1.38-1.51 \AA\ for BPRs. 

\begin{figure}[t]
	\centering
	\includegraphics[scale=0.39,keepaspectratio]{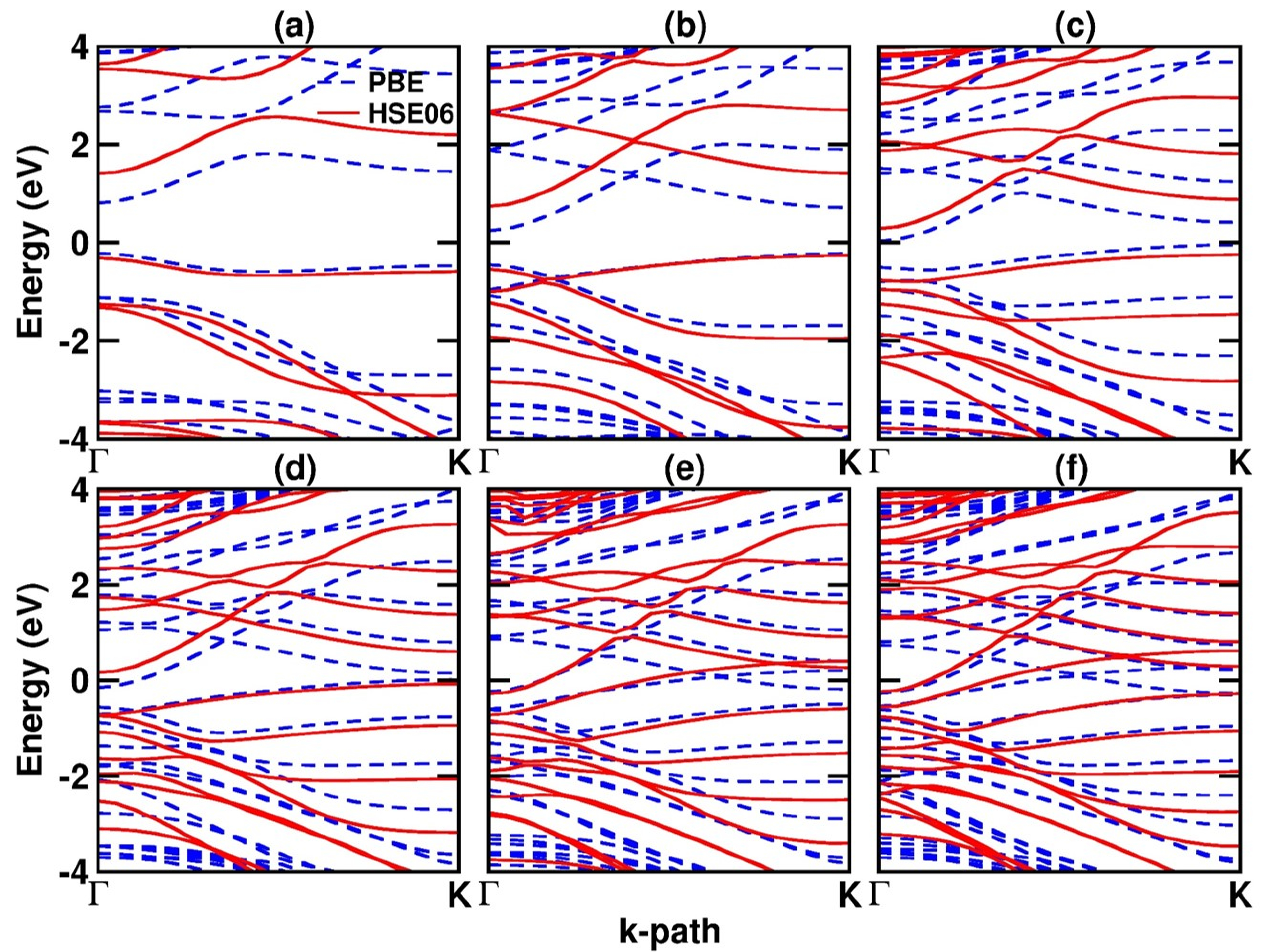}
	\caption{The electronic band structure of (a) 6-BPR, (b) 9-BPR, (c) 12-BPR, (d) 15-BPR, (e) 18-BPR, and (f) 21-BPR, calculated using PBE (dashed blue line) and HSE06  (solid red line) functionals. The Fermi level (E$_F$) is set at 0 eV.} 
	\label{Fig. 2}
\end{figure}

\begin{figure*}[t]
	\centering
	\includegraphics[scale=0.65,keepaspectratio]{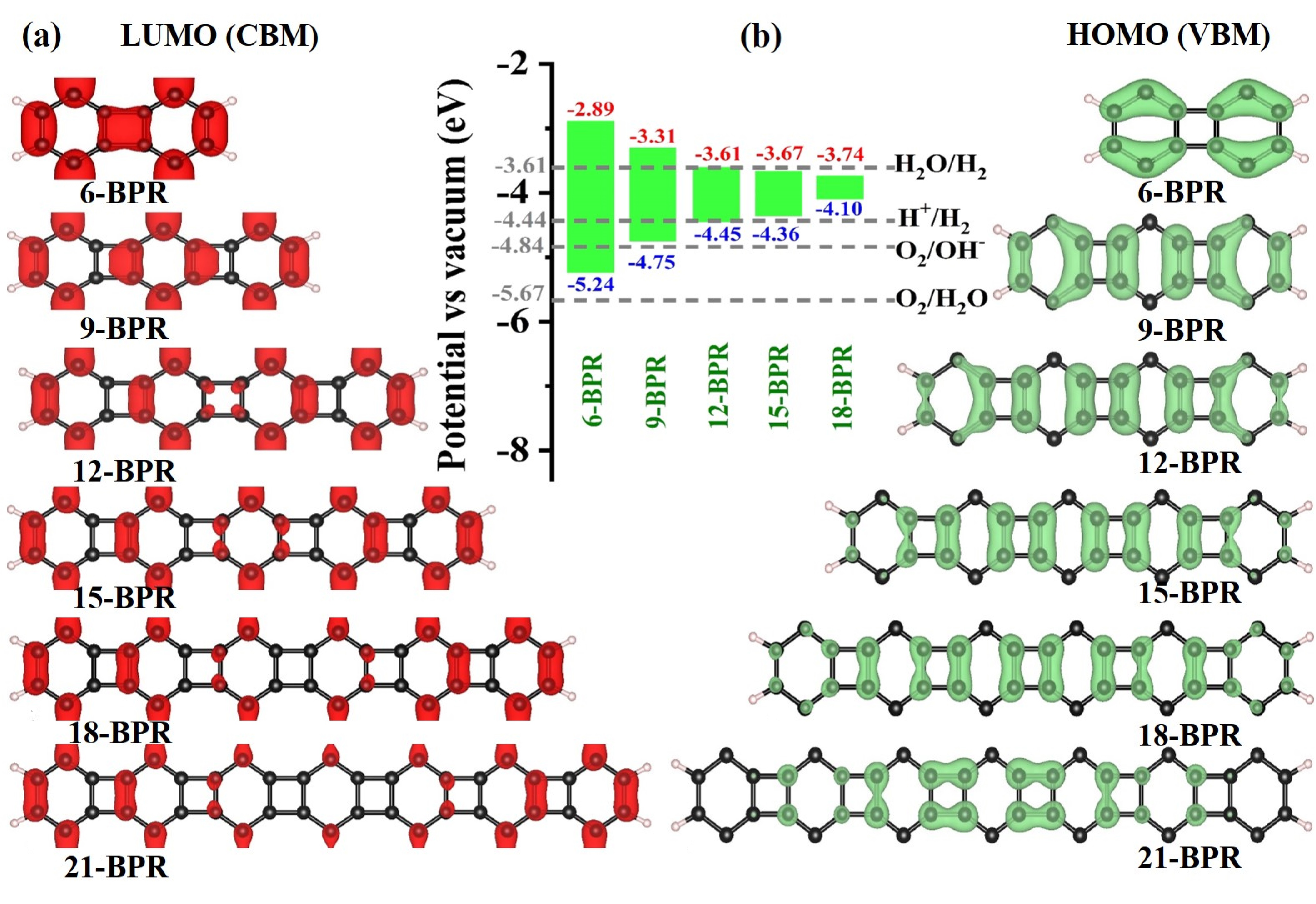}
	\caption{(a) Partial band decomposed charge density for all N-BPRs. Isosurfaces indicating the lowest unoccupied molecular orbital (LUMO) and highest occupied molecular orbital (HOMO) are highlighted by red (left) and green (right) color with isovalue of 0.003 e/\AA$^3$, respectively. (b) Band edge alignments corresponding to the VBM and CBM energies with respect to vacuum against the water redox potentials calculated using PBE functional but are scissor shifted to the experimental band gap. C (H) atoms are represented by black (pink) spheres.}  
	\label{Fig. 3}
\end{figure*}

	Figure ~\ref{Fig. 2} shows the electronic band structure computed using PBE and HSE06 functional for N-BPRs. Only 6-BPR exhibits direct band gap characteristics at the $\Gamma$-point. While, for 9-, 12-, and 15-BPR acquire indirect band gap with valence band maxima (VBM) located at K-point and conduction band minima (CBM) at $\Gamma$-point. For 18- and 21-BPR, metallic characteristics are observed. The computed values of energy band gaps at HSE06 level are 1.62, 1.00, 0.42, and 0.25 eV for 6-, 9-, 12-, and 15-BPR, respectively. The band gaps are observed to decrease as the nanoribbon gets wider, eventually becoming metallic at 18-BPR within the HSE06 calculation. This trend is consistent with the experimental findings. Such behavior is attributed to the quantum confinement of electrons in BPRs. With increasing BPR width, it eventually leads to the 2D Biphenylene metallic character when the width N$\rightarrow$$\infty$\cite{son2006energy,topsakal2009first}.  Next, we investigated the band decomposed charge density of the frontier orbitals contributing at or near the E$_F$, as shown in Fig.~\ref{Fig. 3}(a). For 6-BPR, the charge density of VBM is localized separately on the two-half of the hexagonal rings, whereas, that of CBM is delocalized on the carbon atoms in the periodic direction of BPR. The charge density pattern of CBM remains similar as the width increases, getting more delocalized at the ends of the nanoribbon. Except for 6-BPR, other nanoribbons are indirect in nature with VBM lying at K-point and hence one may observe that the charge density is localized mostly on the tetragonal rings. As the BPR gets wider, the charge density becomes localized on the central tetragonal rings. It is important to note that as one moves from the nanoribbon edges towards the center, the bond distances between the carbon atoms at the tetragonal rings increases. This indicates that the bond strength of center tetragonal carbon atoms becomes weaker as compared to the corner ones resulting in the decrease of electronic band gap with increasing width. Shen \textit{et al.}\cite{shen2022electronic} has also shown that weak coupling of tetragonal rings in BPRs are responsible for shrinking of band gap at the Y-point using COHP analysis. Fan \textit{et al.}\cite{fan2021biphenylene} experimentally synthesized BPRs indicating a band gap of $\sim$2.35 eV for 6-BPR. In addition, they also utilized an atomic orbital based all electron code FHI-AIMS to simulate a band gap of $\sim$1.81 eV. Hudspeth \textit{et al.}\cite{hudspeth2010electronic} computationally studied the narrowest nanoribbon and reported a band gap of 1.71 eV at HSE/6-31G** level using all-electron Gaussian code. Clearly, the theoretical band gaps are smaller compared to the experimentally measured ones, a well-known fact of any DFT method\cite{fan2021biphenylene}. Our calculated band gap for 6-BPR is smaller than previous theoretical reports, which is possibly due to the different method adopted in the present work. However, the decreasing trend of the electronic band gap with increasing BPR width has been accurately identified and is consistent with the experimental observation, also see Fig. S1 in Supporting information\cite{ESI}. 
    From the projected density of states (DOS) (see Fig. S2 in Supporting information\cite{ESI}), it is clear that for all the BPRs, only the p$_z$ orbitals of carbon atoms contribute to the valence and conduction bands. At around -3.0 eV below the E$_F$, the contribution from other orbitals start to show up. One can also observe that the passivated hydrogen atoms in the N-BPRs do not contribute at/near the E$_F$ (see Fig. S3 in Supporting Information)\cite{ESI}.

\begin{table*}[t]%
	\caption{The deformation potential (E$_d$), effective mass (m$^*$), stretching modulus (C$_{1D}$), and carrier mobility ($\mu$$_{1D}$) for electrons ($e$) and holes ($h$).  m$_0$ refers to the rest mass of electron.}
	\begin{center}
		\begin{ruledtabular}
			\begin{tabular}{ccccccccc}
				N-BPR & E$_d(h)$ & E$_d(e)$ & m$^*_h$ & m$^*_e$ & C$_{1D}$ & $\mu$$_{1D(h)}$ & $\mu$$_{1D(e)}$ \\ [0.5ex] 
                & (eV)  & (eV)  & ($\times$10$^{-3}$m$_0$) & ($\times$10$^{-3} $m$_0$) & ($\times$10$^{10}$ eV/m) & ($\times$10$^{4}$ cm$^2$V$^{-1}$s$^{-1}$) & ($\times$10$^{4}$ cm$^2$V$^{-1}$s$^{-1}$) \\ 
                 \hline\\
				6-BPR & -7.95 & -2.20 & 9.72 & 4.76 & 1.09 & 0.01 & 0.55 \\ [0.5ex]
				9-BPR & 1.45 & -2.10 & 46.69 & 4.24 & 1.59 & 0.06 & 1.04 \\ [0.5ex]
				12-BPR & 1.26 & -2.28 & 48.76 & 4.33 & 2.06 & 0.10 & 1.10 \\ [0.5ex]
				15-BPR & 1.62 & -1.71 & 49.58 & 4.32 & 2.63 & 0.07 & 2.51 \\ [0.5ex]
				18-BPR & 1.20 & -1.23 & 7.08 & 4.56 & 3.05 & 2.80 & 5.16 \\ [0.5ex]				
				21-BPR & -1.44 & -1.18 & 5.39 & 4.55 & 3.39 & 3.27 & 6.29 \\ [0.5ex]				
			\end{tabular}
		\end{ruledtabular}
	\end{center}
	\label{table:table2}
\end{table*}

	Carrier mobility is another important descriptor which play a crucial role in dictating the efficacy of a candidate material for photovoltaic/photocatalysis applications along with the energy gap and effective charge carrier separation. The carrier mobility ($\mu_{1D}$) is calculated considering the effective mass approximation within the deformation potential (DP) theory proposed by Bardeen and Shockley\cite{bardeen1950deformation}. This takes into account the effect of acoustic phonon scattering in the long-wavelength limit, providing an upper bound on the mobility estimates\cite{fei2014strain}. The anisotropic nature of transport is well captured within this approximation. For 1D materials, the $\mu_{1D}$ for both the charge carriers at temperature $T$ is given by, 	
    \begin{equation}
		\mathrm{\mu_{1D}}= \frac{e\hbar^2 C_{1D}}{\sqrt{2\pi k_B T} \mid m^*\mid^\frac{3}{2} (E_d)^2}
		\tag{1}
    \end{equation} 
	where e, $\hbar$ and k$_B$ represents the charge of electron, reduced Planck constant and Boltzmann constant respectively. The stretching modulus ($C_{1D}$=[$\partial$$^2$E/$\partial$$\delta$$^2$]/$L_0$) is defined as lattice deformation activated by strain ($\delta$). $L_0$ being the lattice constant of nanoribbon. The carrier effective mass of holes or electrons ($m^*$=$\hbar$$^2$/[$\partial$$^2$E/$\partial$k$^2$]) can be extracted from the electronic band structures in the vicinity of VBM or CBM with respect to the high symmetric k-points $\Gamma$-K along the nanoribbon direction. The deformation potential constant (E$_d$=$\partial$E$_{edge}$/$\partial$$\delta$) indicates the scattering of holes or electrons by the acoustic phonon across the periodic direction.

    Table~\ref{table:table2} presents the computed values of deformation potential (E$_d$), effective mass (m$^*$), stiffness constant (C$_{1D}$),  and carrier mobility ($\mu_{1D}$) for both charge carriers of all N-BPRs (N=6, 9, 12, 15, 18, 21). The BPR width has significant impact on the calculated C$_{1D}$ which increases with increase in the number of carbon atoms (also see Fig. S4 in Supporting information\cite{ESI}). This confirms that higher energies are required to compress/elongate more C-C bonds in N-BPRs (enhancement in the rigidity). Next, we calculate E$_d$ which defines the scattering of an electron (CB edge) or hole (VB edge) by the acoustic phonon. Under the low carrier concentration condition, the charges reside near the band edges. So, the energy change at the upper (lower) edge of valence (conduction) band for holes (electrons) is considered. Slope of a linear fit of these band edges as a function of strain gives the DP constant (see Fig. S5 in Supporting information\cite{ESI}). The scattering of holes or electrons can be well understood by examining the frontier orbitals responsible for the carrier transport. The highest occupied molecular orbital (HOMO) of 6-BPR has localized charges along the length of nanoribbon and hence the energy change will be smaller (see Fig.~\ref{Fig. 3}(a)). Similar analogy holds true for 21-BPR also, where one can see even more localized charges around the central tetragonal rings as compared to the ends. From 9- to 18-BPR, there are localized charges along the periodic direction of nanoribbon and hence is more prone to energy changes. On the contrary, lowest unoccupied molecular orbital (LUMO) has delocalized charges along the periodic direction of nanoribbon. Delocalized ones are less prone to energy changes and hence have smaller E$_d$ constant. The extended delocalized orbitals will be less sensitive to the bond distance changes as compared to the localized ones.
    The effective mass of electron (m$^*_e$) at the CBM at $\Gamma$-point for all BPRs exhibits similar dispersion and hence yield similar values of m$^*_e$. On the other hand, the effective mass of hole (m$^*_h$) at the VBM at $\Gamma$-point for 6-BPR is a little less dispersive and hence a higher (almost double of electrons) m$^*_h$ is observed. For 9-, 12-, and 15-BPR, a flat nature of valence band is observed along the $\Gamma$-K direction, giving rise to very heavy hole (m$^*_h$) as compared to the electron mass.  As we move to wider nanoribbons, \textit{i.e.} 18- and 21-BPR, the top valence band do not retain flat band characteristics. Figure S6 of Supporting information\cite{ESI} indicates a closer view of the topology of bands near the E$_F$ where the valence band inside ellipse deviate from its flat nature. Higher values of m$^*$ for holes compared to electrons suggests higher carrier transport for electrons.

\begin{figure*}[t]
	\centering
	\includegraphics[scale=0.7,keepaspectratio]{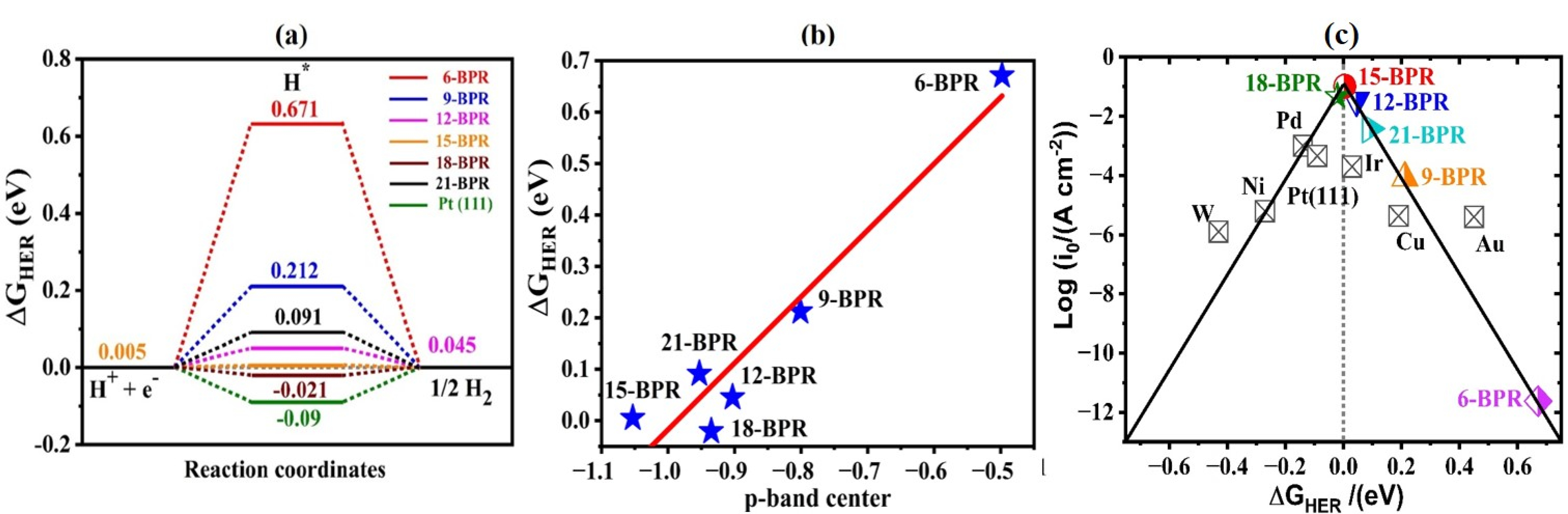}
	\caption{(a) Gibbs free energy profile, and (b) the corresponding p-band centers for HER activity of various N-BPRs. Results for state-of-the-art Pt(111) is given for comparison. (c) HER volcano curve of the exchange current density (log i$_0$) vs. $\Delta$G$_{HER}$ for various N-BPRs, compared with the experimental data of various pure metals and state-of-the-art Pt(111), taken from Ref.\cite{norskov2005trends}.} 
	\label{Fig. 4}
\end{figure*}

    For N-BPRs (N=6-21), the electron mobility lies in the range 0.55$\times$10$^4$ to 6.29$\times$10$^4$ cm$^2$ V$^{-1}$s$^{-1}$, while the hole mobility varies between 0.01$\times$10$^4$ and 3.27$\times$10$^4$ cm$^2$ V$^{-1}$s$^{-1}$. Both electron and hole mobility increases with increase in the width of BPR. $\mu_{1D}$ for electrons is much higher as compared to holes for all N-BPRs indicating n-type electronic characteristics. The relatively lower $\mu_{1D}$ for holes is mainly attributed to flat valence bands around the K-point. For semiconducting BPRs, the electron mobility is around two orders of magnitude higher than the hole mobility. But for metallic BPRs, the electron mobility is almost double than that of hole mobility. These mobility values are of the same order as that of Graphdiyne nanoribbons (2$\times$10$^4$  to 34$\times$10$^4$ cm$^2$ V$^{-1}$s$^{-1}$)\cite{long2011electronic}, graphene nanoribbons (4.09$\times$10$^4$ cm$^2$ V$^{-1}$s$^{-1}$)\cite{wang2012effect}, single-walled carbon nanotubes (7.9$\times$10$^4$ cm$^2$ V$^{-1}$s$^{-1}$)\cite{durkop2004extraordinary}. Apart from this, the partial band decomposed charge density also corroborates well with our findings. Figure ~\ref{Fig. 3}(a) indicates that the charges residing at both the VBM and CBM are spatially separated. The charges associated with VBM mainly resides on central tetragonal rings along the width of N-BPR, whereas those for CBM resides on the hexagonal carbon atoms at the edges along the periodic direction. These spatially separated carriers may help in effective separation of both the charge carriers with low chances of electron-hole recombination.

	To evaluate the performance of a catalyst for water splitting reactions, it is necessary to have band alignments straddling the water redox potentials\cite{tan2015computational,wong2018enhancing}. Figure~\ref{Fig. 3}(b) shows the band edge alignments for the N-BPRs referenced with respect to vacuum against the water redox potentials. In alkaline conditions (H$_2$O/H$_2$ and OH/H$_2$O), 6-BPR is capable of splitting water as a single material as its band alignments straddles both reactions. However, in acidic conditions (H$^+$/H$_2$ and O$_2$/H$_2$O), all the BPRs overcomes only the reduction side indicating its applicability for HER application alone. We further investigated the HER activity of N-BPRs, which is a two-electron reduction process described as H$^+$+2e$^-$$\rightarrow$H$_2$. The Gibbs free adsorption energy for the H adsorbed on nanoribbons in acidic media (\textit{p}H=0) is calculated using the following equation\cite{man2011universality}, 
 \begin{equation}
		\tag{2}
		\Delta \mathrm G_{H^*} = \Delta \mathrm E_{H^*} + \Delta \mathrm E_\mathrm {ZPE}-\mathrm T\Delta \mathrm S_{H} + \Delta \mathrm G_{ \textit{p}\mathrm H} + \Delta \mathrm G_{\mathrm U} + \Delta \mathrm{G_{\mathrm{sol}}}
	\end{equation}
    Here, the first term on the right represents the adsorption energy of hydrogen, defined as $\Delta$E$_{H^*}$= $\Delta$E$_{(*+cat)}$-$\Delta$E$_{cat}$-(1/2)E$_{H_2}$. $\Delta$E$_{(*+cat)}$ and $\Delta$E$_{cat}$ represents the total energies of BPRs with and without H-adsorption, respectively. The second and third term represents the zero point energy and entropy differences between the adsorbed H-atom and H$_2$ gas phase. The fourth and fifth term involves \textit{p}H dependence and external potential bias (U=0), respectively. $T$ is the temperature taken as 300 K. The entropy of H$_2$ gas molecules has been derived from the NIST database\cite{johnson2013nist}. Additionally, to account for solvation energy effects ($\Delta$G$_{sol}$), we have applied continuum solvation model with the dielectric constant for water solvent taken as 78.4\cite{VASPsol-Software,mathew2014implicit}.  Strong adsorption energy leads to difficulties in H desorption, while weak adsorption leads to difficulties in attracting protons to the catalyst surface. As such, the Gibbs free energy should neither be too strong nor too weak. In accordance with Sabatier principle and to achieve maximum reaction rate, the Gibbs free energy for H adsorption should be close to zero. 
 
    We modelled all the available adsorption sites on a 211 supercell of N-BPRs. There are five adsorption sites including (1) the top of C1 carbon atom (2) the top of C2 carbon atom (3) center of tetragonal carbon ring (4) center of hexagonal carbon ring and (5) center of octagonal carbon ring, as illustrated in Fig. S7 of Supporting information\cite{ESI}. Among all, C2 site is found to be energetically the most favourable for all the nanoribbons at a height of 1.11 \AA. Bader charge analysis reveals that the H-atom loses an overall average charge of 0.084, 0.083, 0.072, 0.068, 0.070, and 0.043 e$^-$ to the 6-BPR, 9-BPR, 12-BPR, 15-BPR, 18-BPR, and 21-BPR, respectively. Figure~\ref{Fig. 4}(a) indicates the calculated Gibbs free energy for HER ($\Delta$G$_{HER}$) for the nanoribbons. In terms of magnitude, it is observed that $\Delta$G$_{HER}$ first decreases as we go from 6-BPR to 15-BPR and then increases for 21-BPR. The calculated values of $\Delta$G$_{HER}$ are 0.671, 0.212, 0.045, 0.005, -0.021, and 0.091 eV for 6-BPR, 9-BPR, 12-BPR, 15-BPR, 18-BPR, and 21-BPR, respectively. Among these, 12-BPR to 21-BPR are found to be better HER catalysts with Gibbs free energy in the range -0.2 to 0.2 eV. Most importantly, 15-BPR holds the smallest value of $\Delta$G$_{HER}$, much smaller compared to the state-of-the-art Pt(111) (-0.09 eV). 

    \begin{figure*}[t]
	\centering
	\includegraphics[height= 12 cm, width=17.5 cm]{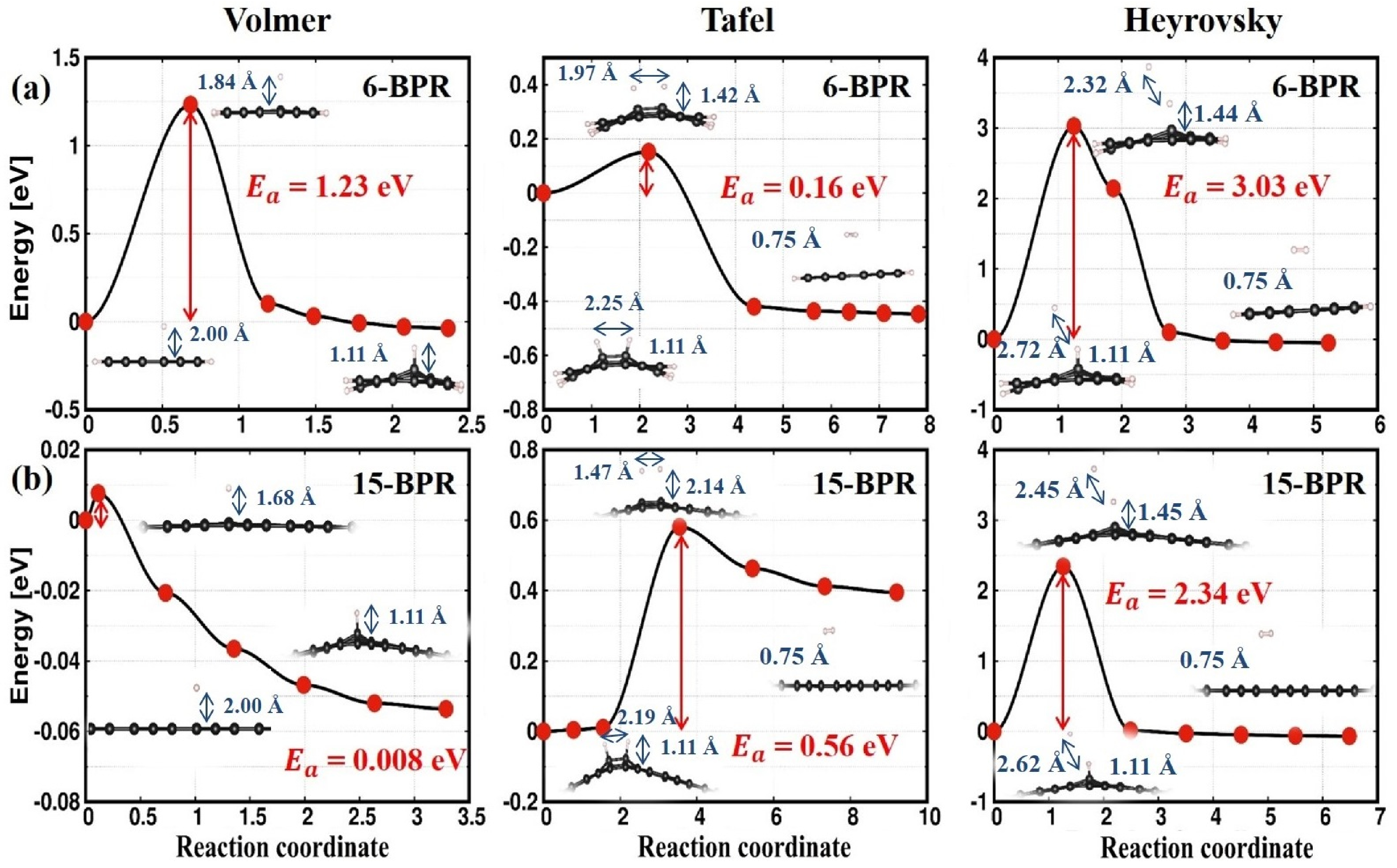}
	\caption{(a) Kinetic barriers of the minimum energy pathway for the Volmer (left), Tafel (middle), and Heyrovsky reactions (right) for (a) 6-BPR and (b) 15-BPR. The insets show the inital state (IS), transition state (TS), and the final state (FS). E$_a$ represents the activation barrier.} 
	\label{Fig. 5}
    \end{figure*}

    To understand the origin of catalytic activity, we estimated the p-band center ($\epsilon_{p}$) of the active site (which can be directly correlated with the HER activity) using the following equation\cite{jacobs2019assessing}, 	
    \begin{equation}
		\tag{3}
		\epsilon_{p} = \frac{\int_{-\infty}^{\infty}\! {E . \rho (E) dE}}{\int_{-\infty}^{\infty} { \rho (E) dE} }
    \end{equation}
	where, $\rho(E)$ is the PDOS of p-orbitals at a given energy (E). The integration was carried out considering only the occupied orbitals in the energy range -0.5 to 0 eV. The states available near the Fermi level (E$_F$) play an important role in determining the catalytic activities. A lower value of $\epsilon_{p}$ with respect to E$_F$ implies strong interaction between the catalyst and the adsorbed H-atom giving small values of $\Delta$G$_{HER}$. In other words, the more positive the $\epsilon_{p}$ relative to E$_F$, the more positive will be the $\Delta$G$_{HER}$. In Fig.~\ref{Fig. 4}(b), the relation between the $\Delta$G$_{HER}$ and $\epsilon_{p}$ is established. It is evident that the $\epsilon_{p}$ for 15-BPR lies far away from the E$_F$, thus leading to small value of $\Delta$G$_{HER}$. A similar trend has been reported before\cite{somaiya2023quasi,sahoo2023activation}. We further investigated the exchange current density, following the Norskov's assumption\cite{norskov2005trends} under standard conditions (\textit{p}H=0 and T=300 K), as defined below,\\ 	
    For $\Delta$G$_{HER}$ $<$ 0 (exothermic process),
    \begin{equation}
	\tag{4}
	\mathrm{i{_0}}=\frac{-e\textit{k$_0$}}{1+exp(-\Delta G_{HER}/{\textit{k}_BT})}
    \end{equation}
    For $\Delta$G$_{HER}$ $>$ 0 (endothermic process),
    \begin{equation}
        \tag{5}
	\mathrm{i{_0}}=\frac{-e\textit{k$_0$} \ exp(-\Delta G_{HER}/{\textit{k}_BT})}{1+exp(-\Delta G_{HER}/{\textit{k}_BT})}
    \end{equation}		
    where, \textit{k$_0$} is the rate constant and \textit{k}$_B$ is the Boltzmann constant.  $\mathrm{i}_0$ is an experimental parameter which evaluates the intrinsic activity of a catalyst to confirm the HER performance. This can be analyzed using a volcano plot depicting the relation between exchange current density and the Gibbs free energy, as shown in Fig.~\ref{Fig. 4}(c). For excellent HER activity, a catalyst requires maximum exchange current density together with $\Delta$G$_{HER}$ close to zero. It is clear from the volcano plot that BPRs with a width from N=12 to N=21 exhibits the highest exchange current density together with low value of $\Delta$G$_{HER}$. Among all, 15-BPR is an outstanding HER catalyst as it possess a free energy close to zero along with maximum exchange current density.

	Lastly, we now discuss the kinetics of two-electron electrochemical HER mechanism. The first step involves the adsorption of an H atom known as the Volmer reaction\cite{santos2011model}. The second step involves the desorption of molecular hydrogen (H$_2$) which can take place either via the Tafel or Heyrovsky reaction\cite{santos2011model}. The former consists of two adsorbed H atoms which combine to form H$_2$, whereas the latter has an adsorbed H atom which reacts with another water molecule to form H$_2$. In acidic media, the two-electron HER process takes place through two competing mechanisms. These are the so-called Volmer-Tafel and Volmer-Heyrovsky mechanisms and are described by the following elementary reactions-
	\begin{equation}
		\tag{R1}
		* + H^+ + e^- \rightarrow H^* (\text{Volmer reaction})
	\end{equation}
	\begin{equation}
		\tag{R2}
		H^* + H^* \rightarrow H_2 (\text{Tafel reaction})
	\end{equation}
	\begin{equation}
		\tag{R3}
		H^* + H^+ + e^- \rightarrow H_2 (\text{Heyrovsky reaction})
	\end{equation}
	where, * indicates the active site on the catalyst surface. Out of the two, one of the mechanism dominates depending upon the catalyst and dictates the rate of the reaction. The transition state nudge elastic band (NEB) calculations are performed within the framework of DFT in conjunction with the climbing image NEB (CI-NEB) which drives the image to the saddle point having the highest energy\cite{henkelman2000climbing,henkelman2000improved}. Activation energy (E$_a$) and the reaction energy ($\Delta$E$_{H^*}$) are two important quantities associated with the transition path connecting the reactants (initial state, IS), the transition state (an intermediate state with higher energy, TS) and the products (final state, FS). The former is defined as the difference between the IS and the transition state (TS), whereas the latter is defined as the difference between the IS and FS.
	
	Here, we have considered the smallest 6-BPR and the best performing 15-BPR as the representatives to understand the HER kinetics, as the former exhibits maximum while the latter exhibits minimum HER overpotential. Figure~\ref{Fig. 5} depicts the minimum energy pathway (MEP) for the Volmer, Tafel and Heyrovsky reactions indicating the IS, TS, and FS  with their corresponding activation barriers.  The most stable site for H adsorption on the surface of BPRs is determined by simulating free energy for all the possible sites. The site with the lowest value of free energy (on top of C2 atom) is considered as the active site for the Volmer reaction on BPRs. For Volmer reaction, the initial C-H bond distance is 2 \AA\ which on structural optimization reduces to 1.11 \AA\ in case of both 6- and 15-BPR. In the TS, the C-H bond distance reduces to 1.84 \AA\ and 1.68 \AA\ for 6-BPR and 15-BPR, respectively. As stated, a more positive value of E$_a$ would mean a higher activation barrier. It is therefore observed that 6-BPR indicates a higher E$_a$ barrier of 1.23 eV as compared to 15-BPR which indicates an E$_a$ of 8 meV.
	
	Next, we model the desorption of H$_2$ following the Tafel reaction. Here, we have assumed that both the H-atoms are bonded to the nearest active C2 atoms. For 6-BPR, the initial C-H and H-H bond distances are 1.11 \AA\ and 2.25 \AA\, respectively. In the TS, the C-H (H-H) bond distance is found to increase (decrease) to 1.42 (1.97) \AA. In the FS, it then evolves out as H$_2$ molecule with an equilibrium H-H bond distance of 0.75 \AA. The calculated minimum energy pathway indicates a barrier of 0.16 eV which is smaller compared to Volmer reaction. Similarly for 15-BPR, an activation barrier of 0.56 eV is obtained which is much higher compared to Volmer reaction. Next, we study the H$_2$ desorption following the Heyrovsky reaction. Here we allowed H$^+$ to interact with adsorbed H atom by placing them more than 2.5 \AA\ apart. For 6-BPR, the initial H-H (C-H) bond distance is 2.72 (1.11) \AA\ which reduces (increases) to 2.32 (1.44) \AA\ in the TS before evolving out as molecular hydrogen. The calculated E$_a$ is 3.03 eV which is much higher compared to Tafel reaction. Hence 6-BPR follows Volmer-Tafel mechanism with Volmer reaction as the rate limiting step. Similarly for 15-BPR, the calculated E$_a$ is 2.34 eV which is again much higher compared to Tafel reaction. In both the BPRs, the E$_a$ corresponding to the Heyrovsky reaction are found to be unfavorable. This suggests that the predominant HER route in 15-BPR will follow the Volmer-Tafel mechanism with Tafel reaction as the rate limiting step. It is interesting to note that, the state-of-the-art catalyst Pt(111) also follows the Volmer-Tafel mechanism with Tafel reaction as the rate-limiting step with an E$_a$ of 0.85 eV\cite{skulason2010modeling}. On comparing the rate limiting step and the activation barrier, we find that 15-BPR is highly active towards HER than Pt(111) owing to a smaller activation barrier by 0.29 eV.
		
	\section{Conclusions}
	In summary, we have systematically investigated the structural, electronic, carrier transport, and photo(electro)catalytic properties of recently synthesized Biphenylene nanoribbons (N-BPRs) by employing first principles calculations. The electronic band gap shows significant dependence on the width of BPRs and is observed to decrease with increase in the width(N) of BPR eventually becoming metallic at 18-BPR. Interestingly, the p$_z$ orbital of C-atoms located at the central tetragonal rings are mainly responsible for the reduction of band gap as width increases. The simulated room temperature electron carrier mobility is found to be as high as 6.3$\times$10$^4$ cm$^2$V$^{-1}$s$^{-1}$. The band edge alignments confirm to only cover the reduction potential of water, which is investigated further to evaluate its potential for HER activity. Among others, the best catalytic activity is shown by 15-BPR indicating a small overpotential of about 0.005 V and a maximum exchange current density, even better compared to the state-of-the-art Pt(111) catalyst. The kinetics reveal that the HER mechanism for 15-BPR is dominated by the Volmer-Tafel pathway with a Tafel barrier of 0.56 eV. We strongly believe that 15-BPR has the potential to possibly over take the traditional Pt(111), and become the next generation most promising  candidate for HER catalytic activity.

	\section*{Acknowledgements}
	R.N.S. acknowledges IIT Bombay for providing Institute postdoctoral fellowship and computational resources (SpaceTime2) to pursue this work. B.C. acknowledges BARC's supercomputing facility where part of the simulations were carried out.

	\bibliography{references.bib} 
	
\end{document}


\title{Supporting Information for \\ Biphenylene Nanoribbon as Promising Electrocatalyst for Hydrogen Evolution}

\author{Radha N Somaiya}
\affiliation{Materials Modeling Laboratory, Department of Physics, IIT Bombay, Powai, Mumbai 400076, India}

\author{Zicong Marvin Wong}
\affiliation{Institute of High Performance Computing, Agency for Science, Technology and Research, 1 Fusionopolis Way, \#16-16 Connexis, Singapore 138632, Singapore}

\author{Brahmananda Chakraborty}
\affiliation{Homi Bhabha National Institute, Trombay, Mumbai-400085, India}
\affiliation{High Pressure \& Synchrotron Radiation Physics Division, Bhabha Atomic Research Centre, Trombay, Mumbai-400085, India}

\author{Teck Leong Tan}
\affiliation{Institute of High Performance Computing, Agency for Science, Technology and Research, 1 Fusionopolis Way, \#16-16 Connexis, Singapore 138632, Singapore}

\author{Aftab Alam}
\email{aftab@iitb.ac.in}
\affiliation{Materials Modeling Laboratory, Department of Physics, IIT Bombay, Powai, Mumbai 400076, India}
	
\maketitle
	
    \textcolor{black}{Here, we provide further auxiliary information about computational details, experimental and theoretical comparision of electronic band gaps for various BPRs, projected density of states for carbon and hydrogen atoms, the stretching modulus, deformation potential, closer view of VBM and CBM for 18-BPR and 21-BPR, and the available adsorption sites on the surface of BPR to further strengthen our message in the manuscript.}
 
	\section{Computational details} \textcolor{black}{The density of states (DOS) was calculated using the tetrahedron method with Bloch corrections \cite{blochl1994improved} for semiconductors, while for metals gussian smearing is considered. The climbing-image nudge elastic band (CI-NEB) \cite{henkelman2000climbing,henkelman2000improved} calculations are performed to investigate the kinetic barriers and the minimum energy pathway for the HER mechanism. To do this, five transition state images were considered along the reaction pathway and were optimized until the normal forces are smaller than 0.05 eV/\AA.}
    
    \textcolor{black}{ To evaluate the deformation potential (E$_d$) and the stiffness constant (C$_{1D}$), we applied small deformations along the transport direction in the range of $\pm$0.25\%. All atomic positions are fully converged to obtain accurate estimation of the elastic properties. The stiffness constant, C$_{1D}$, is estimated by taking second derivative of the total energy with respect to applied strain  (see Fig. S4).  This is a reminiscent of Hook's law in which parabolic potential energy is derived for macroscopic strings. The effective mass is calculated from the curvature of the band structure. The second derivative of energy \textit{vs. k}-points in the vicinity of  conduction band minima (CBM) and valence band maxima (VBM) are fitted to obtain effective mass (m$^*$) for electrons (\textit{e}) and holes (\textit{h}) respectively.}

	\begin{figure*}[t]
		\centering
		\includegraphics[height=10 cm, width=10.5 cm]{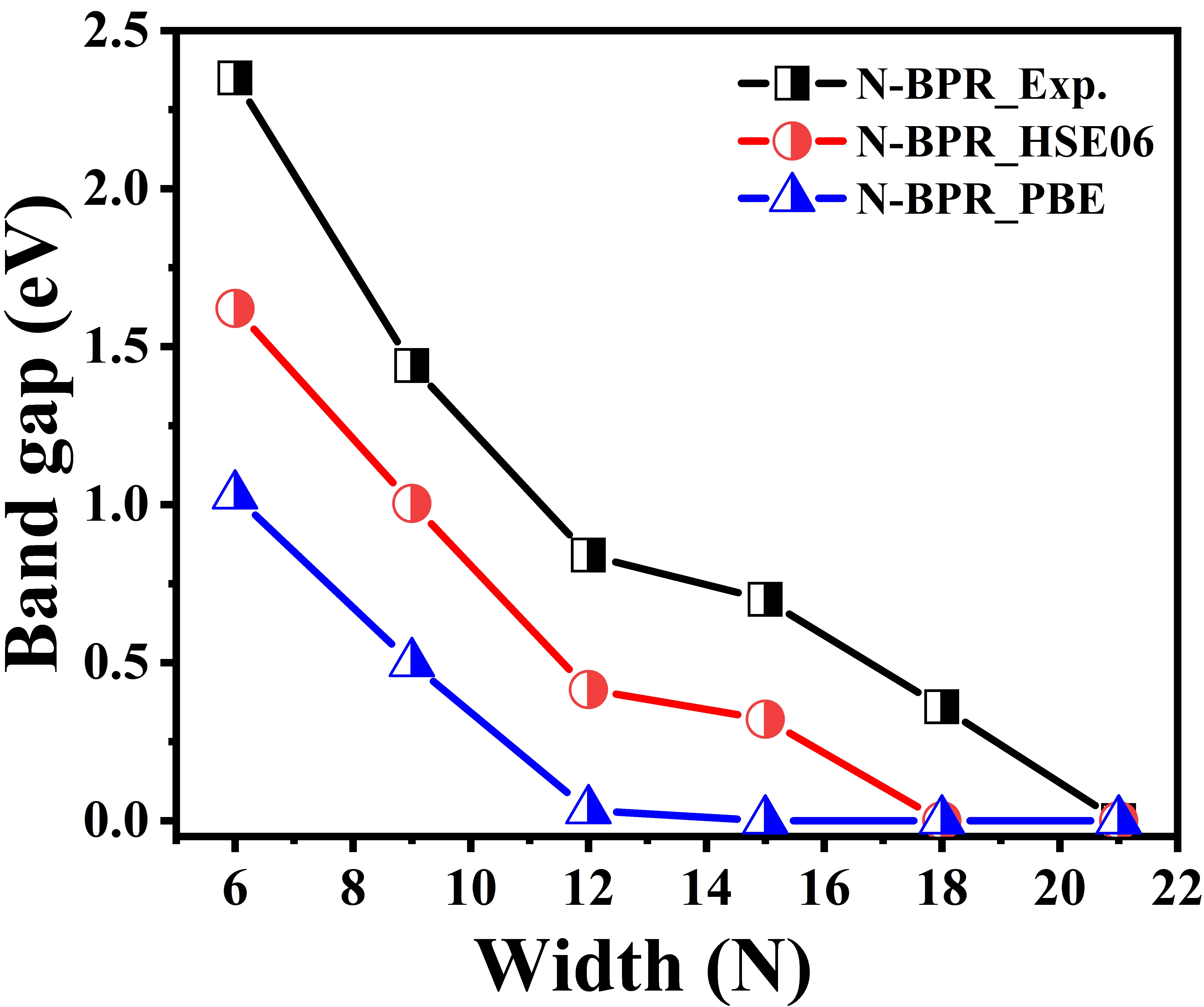}
		\caption{Comparison of simulated electronic band gaps for various BPRs (calculated using PBE and HSE06 functionals) with the experimental values \cite{fan2021biphenylene}. The width is indicated by N, which gives 2N number of atoms in the corresponding nanoribbons.} 
		\label{Fig. S1}
	\end{figure*}
	
	\begin{figure*}[hbt]
		\centering
		\includegraphics[height= 17 cm, width=20 cm]{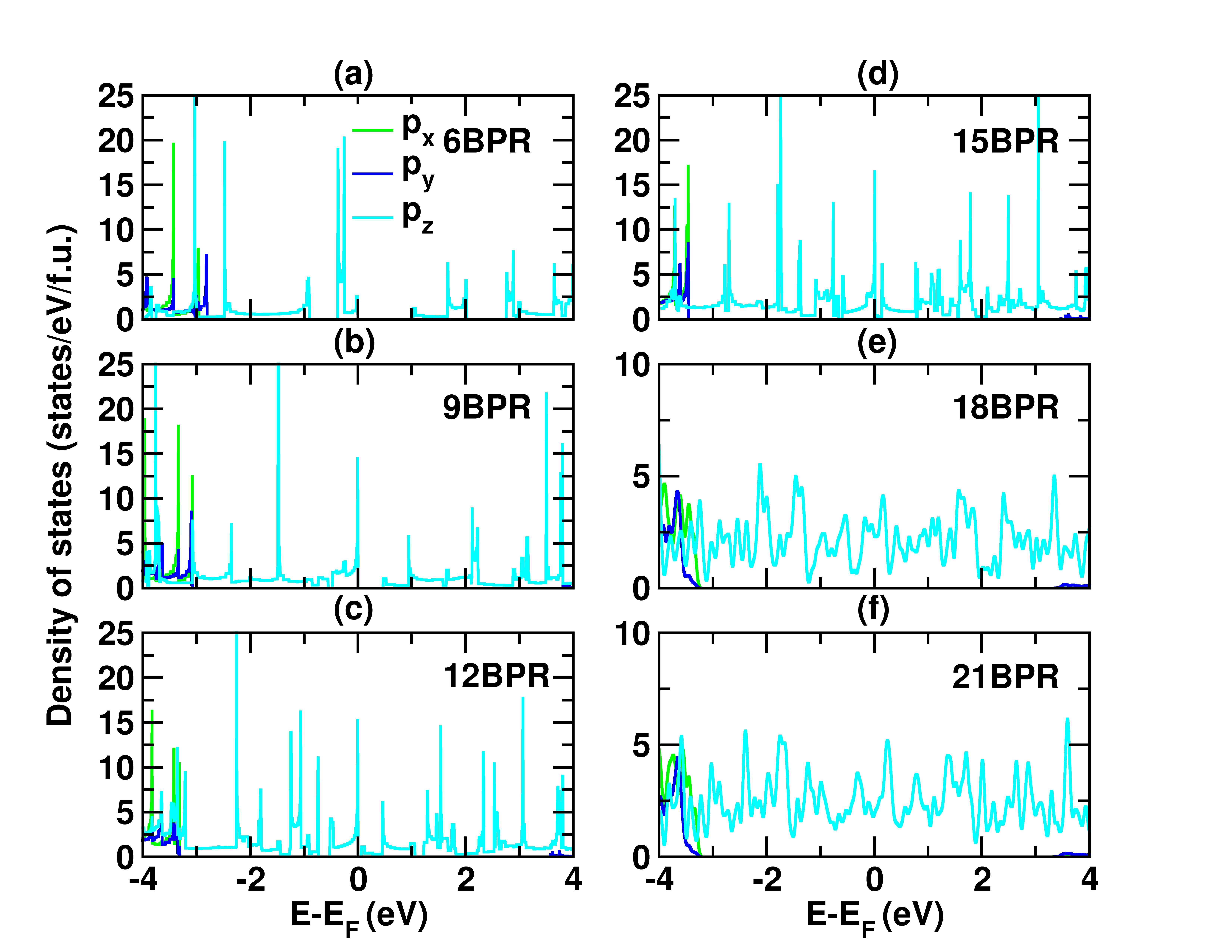}
		\caption{Orbital projected density of states for the carbon atoms for (a) 6-BPR, (b) 9-BPR, (c) 12-BPR, (d) 15-BPR, (e) 18-BPR, and (f) 21-BPR, respectively.} 
		\label{Fig. S2}
	\end{figure*}
	
	\begin{figure*}[hbt]
		\centering
		\includegraphics[height= 17 cm, width=20 cm]{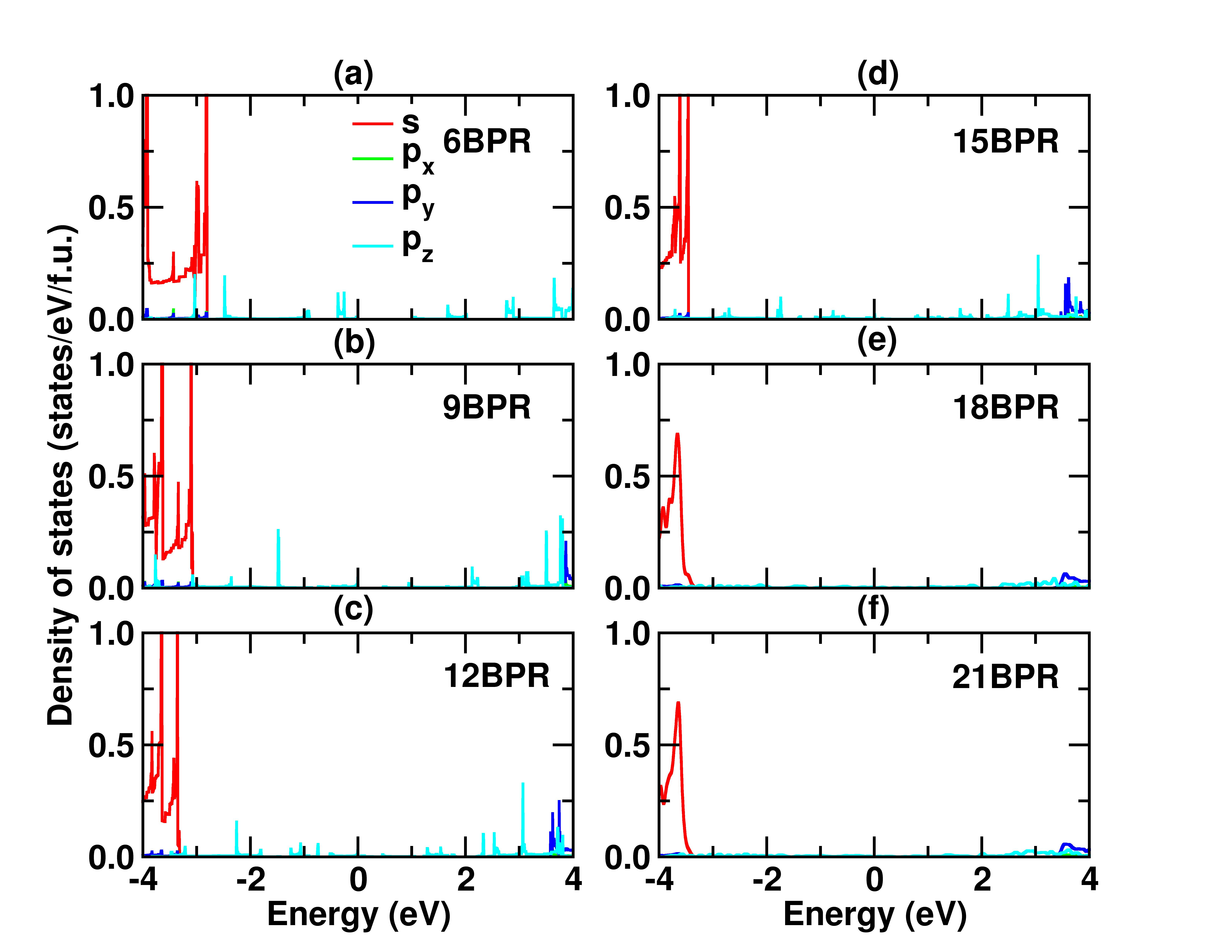}
		\caption{Orbital projected density of states for the passivated hydrogen atoms for (a) 6-BPR, (b) 9-BPR, (c) 12-BPR, (d) 15-BPR, (e) 18-BPR, and (f) 21-BPR, respectively.} 
		\label{Fig. S3}
	\end{figure*}
	
	\begin{figure*}[hbt]
		\centering
		\includegraphics[height= 12cm,width=\columnwidth]{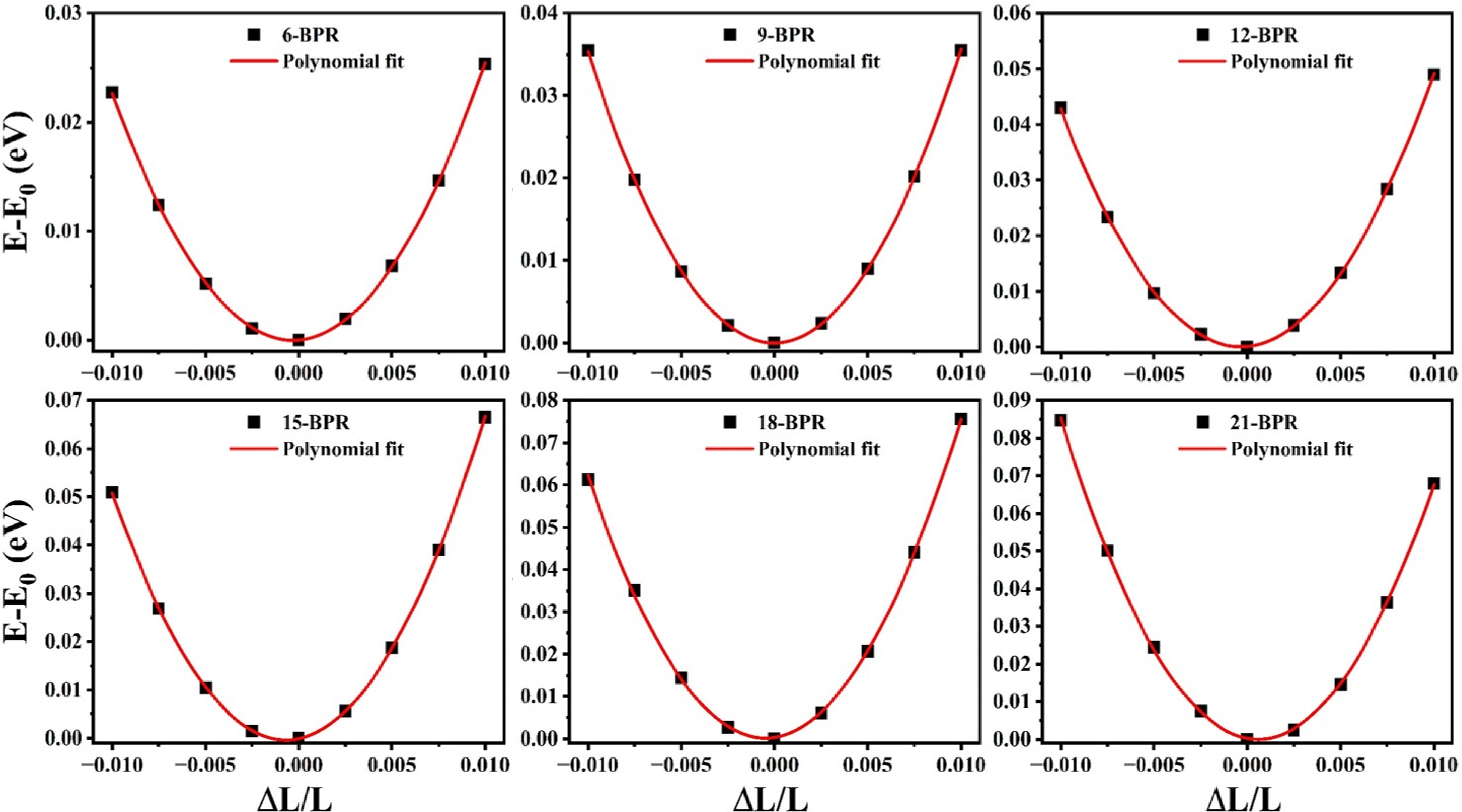}
		\caption{Total energy of N-BPRs (N= 6, 9, 12, 15, 18, 21) as a function of applied strain ($\delta$), for the calculation of stretching modulus (C$_{1D}$).} 
		\label{Fig. S4}
	\end{figure*}
	
	\begin{figure*}[hbt]
		\centering
		\includegraphics[height= 	12cm,width=\columnwidth]{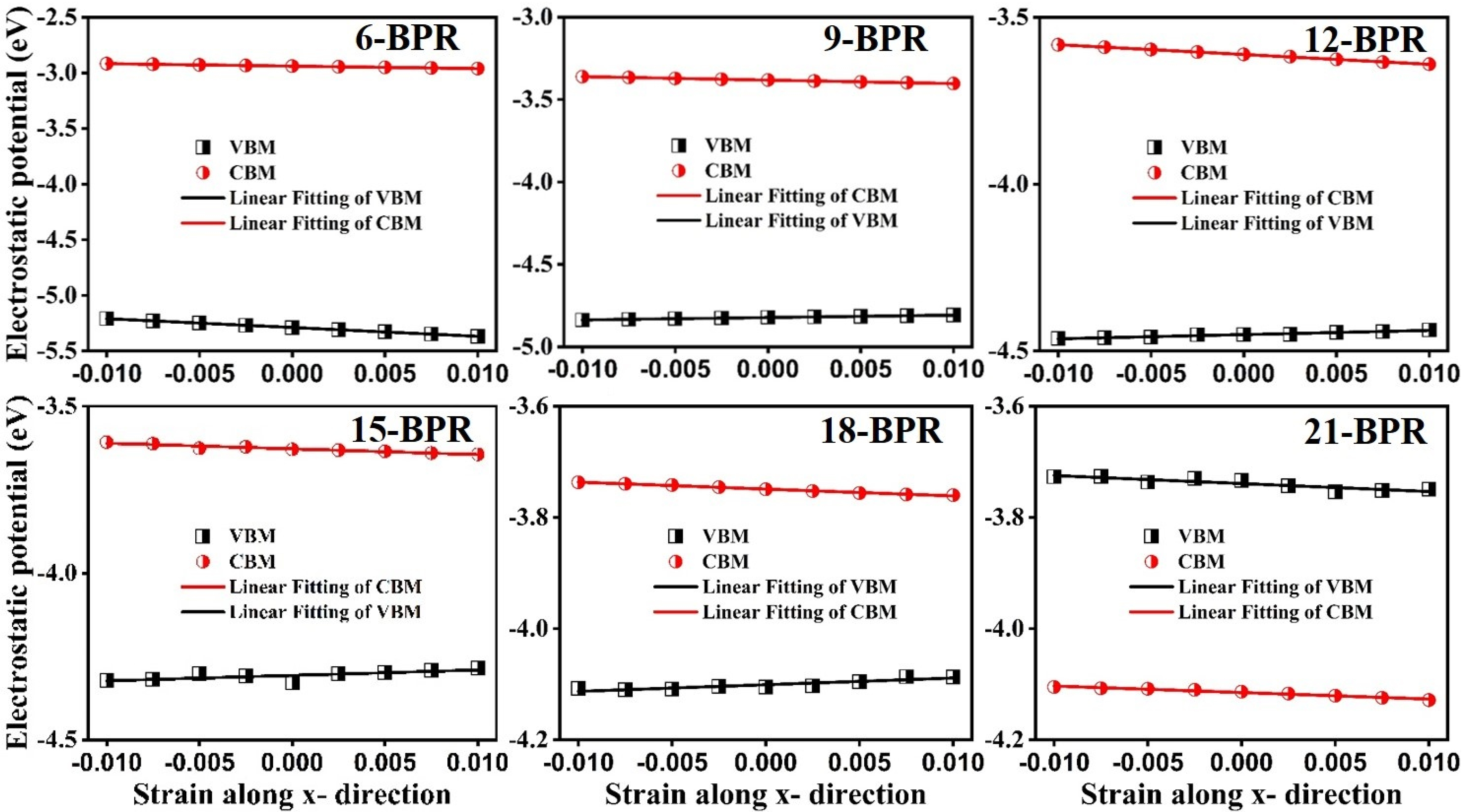}
		\caption{The deformation potential (E$_d$) of various N-BPRs (N= 6, 9, 12, 15, 18, 21) as a function of applied strain ($\delta$), for VBM and CBM.} 
		\label{Fig. S5}
	\end{figure*}	
	
	\begin{figure*}[hbtp]
		\centering
		\includegraphics[scale=0.75,keepaspectratio]{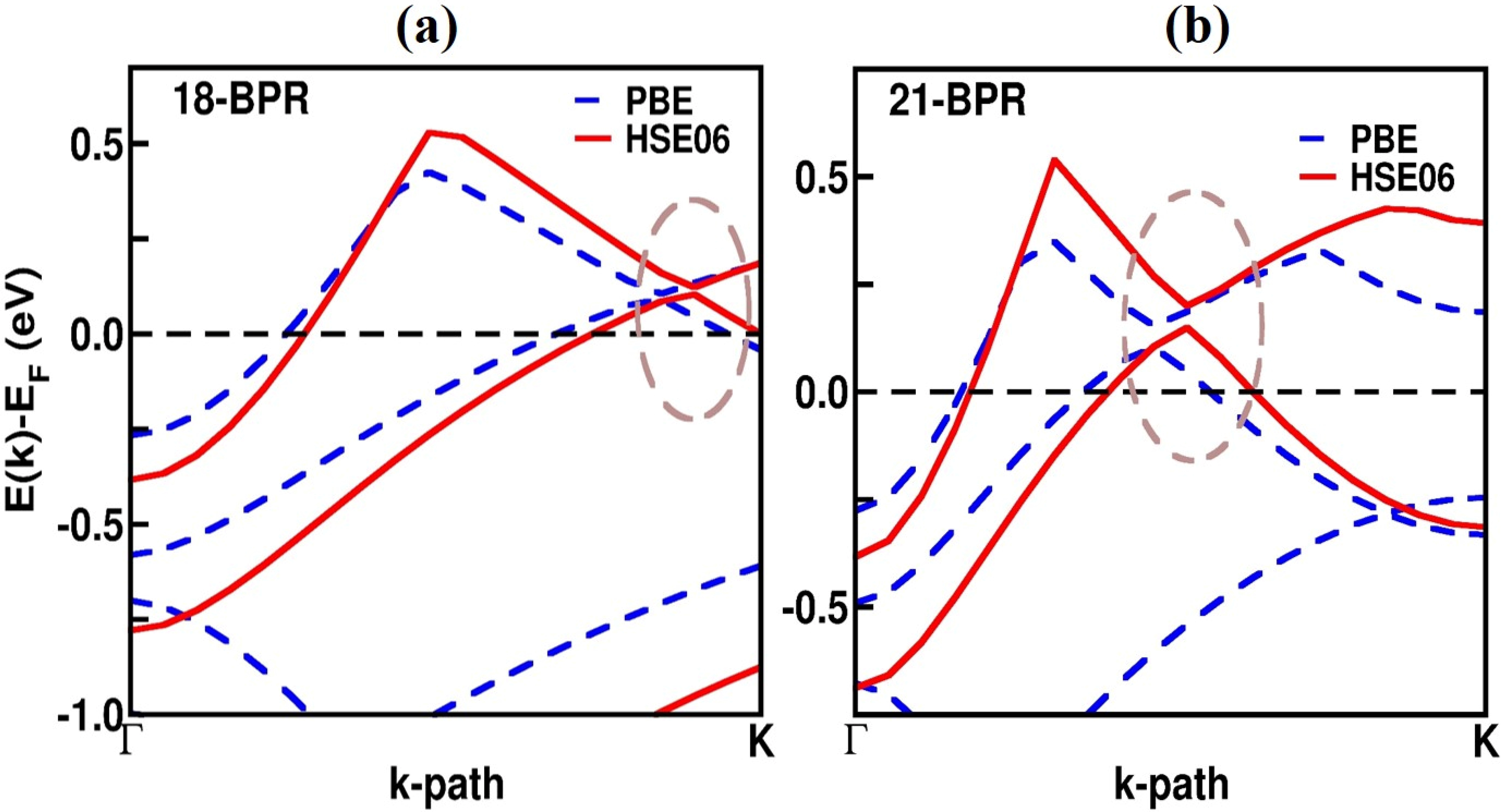}
		\caption{A closer view of the VBM and CBM of (a) 18-BPR and (b) 21-BPR. The brown dotted ellipse indicates the energy range where valence band no longer remains flat.} 
		\label{Fig. S6}
	\end{figure*}	

	\begin{figure*}[hbt]
	\centering
	\includegraphics[scale=0.5,keepaspectratio]{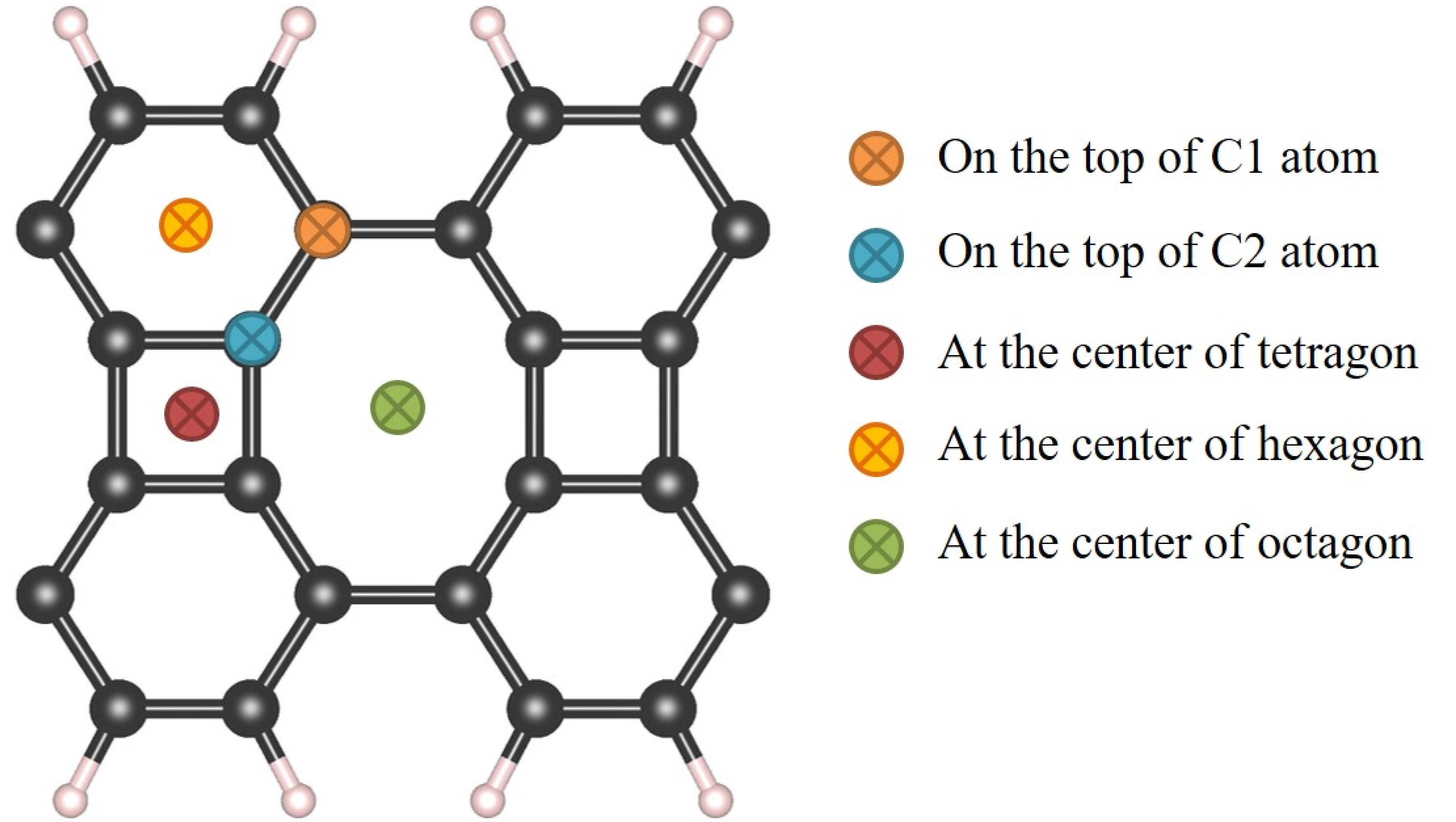}
	\caption{Available adsorption sites on the surface of Biphenylene nanoribbon. C and H atoms are indicated by black and pink spheres, respectively} 
	\label{Fig. S7}
	\end{figure*}

	\FloatBarrier
	
	\bibliography{references.bib}